\documentclass[conference]{IEEEtran}
\IEEEoverridecommandlockouts
\usepackage{cite}
\usepackage{amsmath,amssymb,amsfonts}
\usepackage{algorithmic}
\usepackage{graphicx}
\usepackage{textcomp}
\usepackage{xcolor}
\usepackage{longtable}
\usepackage{algorithm}
\usepackage{amsthm}
\theoremstyle{definition}
\newtheorem{definition}{Definition}
\newtheorem{example}{Example}
\usepackage{array}
\usepackage{ulem} 

\def\BibTeX{{\rm B\kern-.05em{\sc i\kern-.025em b}\kern-.08em
		T\kern-.1667em\lower.7ex\hbox{E}\kern-.125emX}}
\begin{document}
	
	\title{Efficient RDF Graph Storage based on Reinforcement Learning}
	
	\author{
		\IEEEauthorblockN{Lei Zheng}
		\IEEEauthorblockA{\textit{Faculty of Computing} \\
			\textit{Harbin Institute of Technology}\\
			Harbin, China \\
			19S103245@stu.hit.edu.cn}
		\and
		\IEEEauthorblockN{Ziming Shen}
		\IEEEauthorblockA{\textit{Faculty of Computing} \\
			\textit{Harbin Institute of Technology}\\
			Harbin, China \\
			shenziming1120@gmail.com}
		\and
		\IEEEauthorblockN{Hongzhi Wang}
		\IEEEauthorblockA{\textit{Faculty of Computing} \\
			\textit{Harbin Institute of Technology}\\
			Harbin, China \\
			wangzh@hit.edu.cn}
	}
	\maketitle
	
	\begin{abstract}
		Knowledge graph is an important cornerstone of artificial intelligence. The construction and release of large-scale knowledge graphs in various fields pose new challenges to knowledge graph data management. Due to the maturity and stability, relational database is also suitable for RDF data storage. However, the complex structure of RDF graph brings challenges to storage structure design for RDF graph in the relational database. To address the difficult problem, this paper adopts reinforcement learning (RL) to optimize the storage partition method of RDF graph based on the relational database. We transform the graph storage into a Markov decision process, and develop the reinforcement learning algorithm for graph storage design. For effective RL-based storage design, we propose the data feature extraction method of RDF tables and the query rewriting priority policy during model training. The extensive experimental results demonstrate that our approach outperforms existing RDF storage design methods.
	\end{abstract}
	
	
	\section{Introduction}\label{sec:Introduction}
	As the latest achievement of the development of symbolism, the knowledge graph is an important cornerstone of Artificial Intelligence (AI), and it has been continuously developed and applied. In recent years, knowledge graphs with a scale of one million vertices and hundreds of millions of edges have become common in many areas such as Semantic Web~\cite{b1-1}. Many knowledge graphs in the Linking Open Data (LOD) cloud map released in August 2018 exceed 1 billion triples~\cite{b1-2}. For example, both the Wikipedia Knowledge Graph and the Geographic Information Knowledge Graph have more than 3 billion RDF data, and the protein knowledge graph has more than 13 billion RDF data. The construction and release of large-scale knowledge graphs in various fields pose new challenges to knowledge graph management~\cite{b1-3}.
	
	RDF has been widely used in various fields. In the past several decades, RDF plays an important role in query processing. For example, it is used to assist search engines to better find resources, to classify and describe the content and relationships in the websites and libraries, and to represent, share and exchange knowledge in intelligent software agents. There has been a flurry of activity around designing specialized graph analytics engines~\cite{b1-4}. With the increasing applications of RDF, the storage and query requirements of RDF have also put forward higher requirements~\cite{b1-5}.
	
	In the field of knowledge graph management, the relationship-based and the graph-based are two mainstream storage approach directions. Due to the development of the relational database for decades, with a reasonable storage management scheme, it achieves better performance and stability than graph databases for massive data with complex connections~\cite{b1-6}. For this reason, we intend to adopt the relational database to store RDF data.
	
	For relational storage of graph data, it is essential to partition the triple to multiple tables or join some partial results to accelerate query processing. Such determination brings challenges. On the one hand, existing selection methods of graph data storage structures and indices depend on the database administrators (DBAs)~\cite{b1-7}. It is over-demanding for DBAs to decide how to partition and store a large-scale graph~\cite{b1-8}. Besides, it is hardly for human to know the global feature of the complex graph, which causes the non-optimal solution.
	
	On the other hand, due to the rapid update of data and workload in graph data applications, the current database system can not adapt to dynamically changing graph data and workload in real time. When the RDF data or workload changes, the current storage scheme may not still be effective. Dynamic schemas and data sparsity are typical characteristics of RDF data which are not commonly associated with relational databases. Therefore, it is not a coincidence that existing approaches that store RDF data over relational stores~\cite{b1-9,b1-10,b1-11} can not handle this dynamicity without altering their schemas~\cite{b1-12}. These constraints make the graph storage problem difficult to solve. Therefore, there is an urgent need to develop an automatic graph data storage design technique to solve the above problems.
	
	Fortunately, in recent years, reinforcement learning (RL) has been widely used to solve data management problems~\cite{b1-13}. For example, SageDB~\cite{b1-14} uses RL to optimize query scheduling. GALO~\cite{b1-15} proposes the method of guided automated learning for query workload re-optimization. Neo~\cite{b1-16} proposes an end-to-end query plan optimization based on machine learning. SkinnerDB~\cite{b1-17} uses RL to avoid local optimization of the initial query plan. However, there are only a few studies using artificial intelligence for data partition optimization. Schism~\cite{b1-18} uses the partition strategy based on machine learning. GridFormation~\cite{b1-19} tries to use RL methods on the column store database to select data blocks and data distribution methods. Partitioning Advisor~\cite{b1-20} applies the idea of RL to the partition problem of distributed databases.
	
	However, RL can be hardly applied to graph storage design directly. To conduct reinforcement learning on relational storage design for graph data, Markov Decision Process should be defined well on the storage process. It is non-trivial to define state, i.e. how to store complex RDF table sequence into numerical vector. Additionally, since the storage scheme is based on \textsf{dividing} and \textsf{merging} tables, the action space can be infinite. To guarantee as small action space as possible, the action should be defined reasonably. The last but not least, under the changing storage structure, the calculation of reward should be adaptive.
	
	Facing the challenges above, we propose an intelligent storage scheme for graph data based on reinforcement learning. We take predicates as the feature of tables and map the predicates to numerical variables. Besides, we extract features of the workload to restrict the size of the action space. We also propose the priority-based policy for query rewriting to calculate the reward dynamically.	The contributions of this paper are summarized as follows:
	
	\begin{itemize}
		\item We solve the problem of RDF graph storage with reinforcement learning. To the best of our knowledge, this is the first work to adopt RL to solve graph storage problem.
		
		\item We design the effective RL model for the graph storage problem.
		
		\item To verify the performance of the proposed approach, we conduct extensive experiments. Experimental results demonstrate that time performance of our method outperforms Apache Jena.
	\end{itemize}
	
	The remainder of this paper is organized as follows. We start in Section~\ref{sec:Overview} with an overview of our RDF graph storage based on RL. Section~\ref{sec:The Design of Reinforcement Learning for Graph Storage} describes the design of reinforcement learning model for graph storage. In Section~\ref{sec:Query rewrite}, we discuss the query rewrite policy. Section~\ref{sec:Experimental results} presents the experimental results of our system and the comparison with Apache Jena. In the end, Section~\ref{sec:Conclusion} concludes our system and put forward the improvement of future work.
	
	\section{Overview}\label{sec:Overview}
\begin{figure}[htb]
		\includegraphics[width=\linewidth]{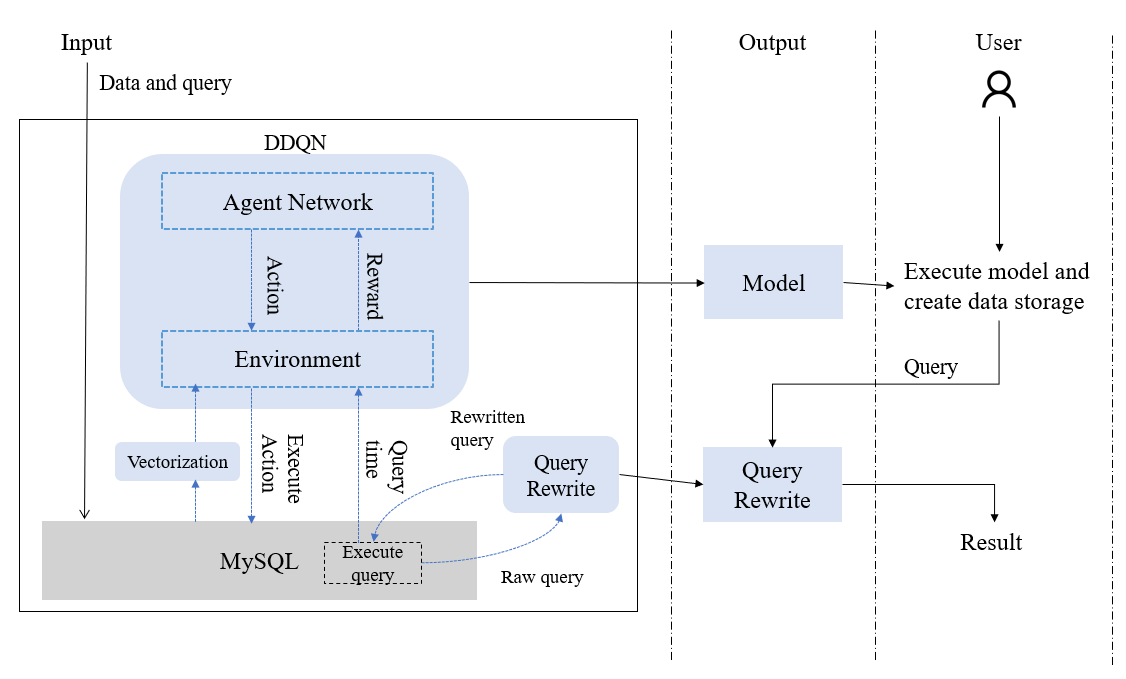}
		\centering
		\caption{Framework}\label{framework}
	\end{figure}

	In this paper, we aim at finding a solution for this challenging problem of RDF graph storage. We realize the graph storage design based on relational database with reinforcement learning. Our idea is to transform the graph storage problem into a Markov decision process (MDP), and define the five factors of the MDP related to graph storage. On this basis, we perform feature extraction on the data and train the reinforcement learning network. At the same time, the query rewrite policy is optimized. The framework of our work is shown in Figure~\ref{framework}.
	
	The system runs as follows: (1) Input module is to input dataset and workload. The dataset is stored in a table in the form of triples. (2) Next, vectorization module extracts and converts the data storage features to a vector. It will be introduced in Section~\ref{subsec:Predicate vector mapping}. (3) DDQN module makes the storage decision by reinforcement learning. It selects the action for the database to execute. And the database returns the workload query time for the calculation of the reward. DDQN continuously interacts with the database environment to update the Q value. This will be introduced in Section~\ref{subsec:Double Deep Q-Network Deep Learning Network}. (4) Query rewrite module rewrites the query to minimize the query execution time under the current storage structure. When the storage structure changes, the query rewrite policy ensures high performance by adapting the query execution strategy to the changed storage. This module will be introduced in Section~\ref{sec:Query rewrite}. (5) After the model is trained, the user can execute the model to establish a data storage structure.
	
	This system can realize the optimal storage scheme of the graph. We extract and vectorize the dataset features as the input of DDQN. DDQN outputs the Q value of a certain action under the current state, and selects the action with the largest Q value to decide whether to \textsf{divide} or to \textsf{merge} tables. Next, the environment executes the rewritten query statement and returns the execution time. The model updates the Q value until the Q value converges or the episode reaches the maximum number of iterations. Thus, the whole system implements automatic data storage.
	
\begin{figure}[htb]
		\includegraphics[width=\linewidth]{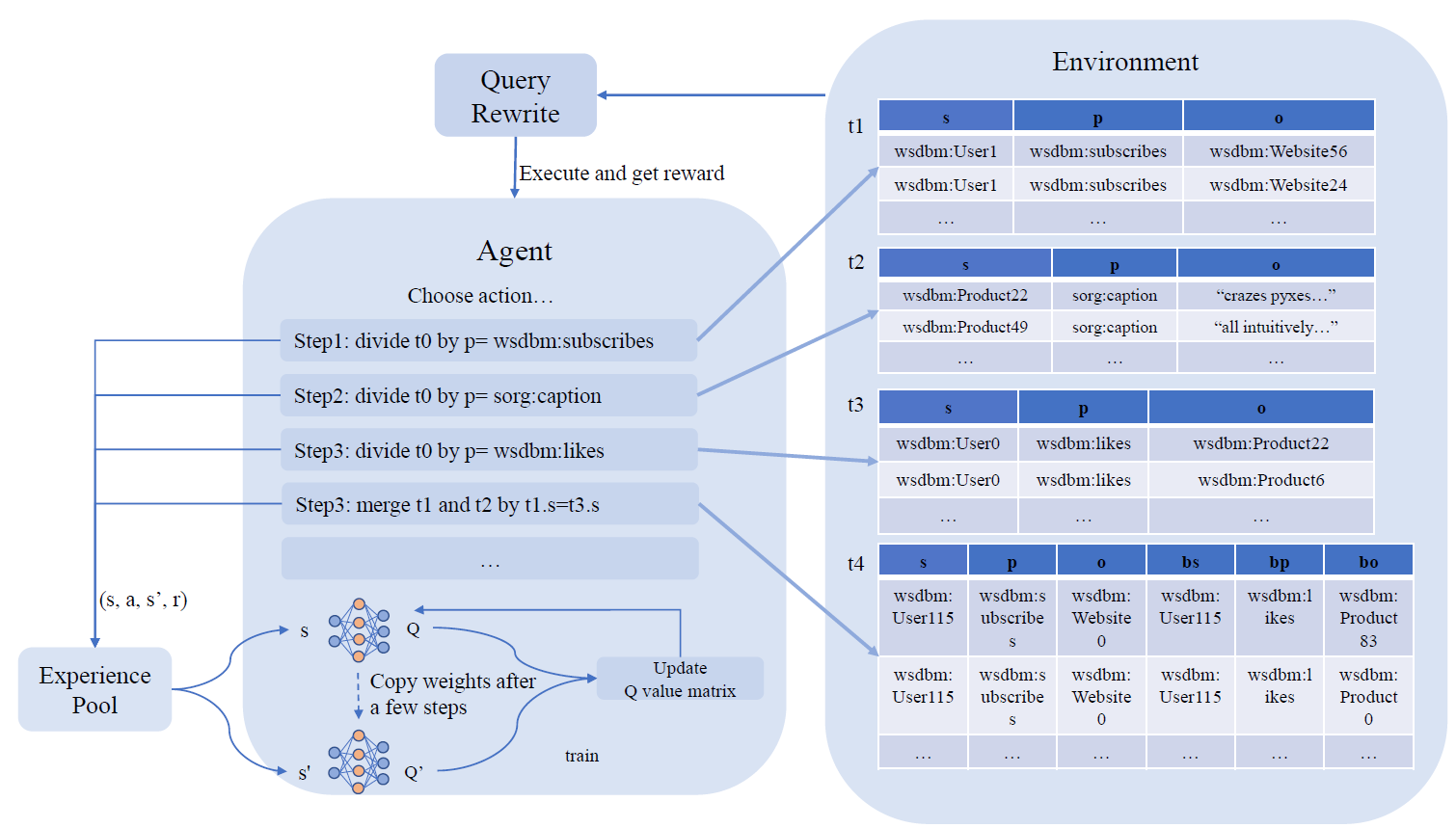}
		\centering
		\caption{Flowchart}\label{Flowchart}
	\end{figure}

	The flow of the agent is shown in Figure~\ref{Flowchart}. The agent chooses actions of \textsf{dividing} or \textsf{merging} tables according to its decision policy. For each step, the quadruples of \textsf{current state}, \textsf{action}, \textsf{next state} and \textsf{reward} are stored in experience pool. After each round, the fixed scale replay batch is used for neural network weights updating. These two networks take the \textsf{current} and \textsf{next state} as the input, respectively, and outputs Q and Q', which are used for updating Q value matrix. Then the Q-evaluation is trained with the current states batch and Q value matrix. After a few rounds, the weights of evaluation-network are copied to the target-network. In environment, storage scheme will change with the agent's actions. The reward of each step is calculated with promotion of the proposed strategy.
	
	\section{The Design of Reinforcement Learning for Graph Storage}\label{sec:The Design of Reinforcement Learning for Graph Storage}
	Reinforcement learning is a continuous decision-making process. Its basic idea is to maximize the cumulative reward value, which is achieved by continuously interacting with the environment and learning the optimal strategy together~\cite{b3-1}.	In this section, we propose a reinforcement learning method to make data storage automatic and intelligent. It uses predicates to characterize table data and adopts DDQN for action selection. By interacting with the database environment, we make the database automatically learn the storage strategy of the data to achieve the purpose of the optimal storage scheme.
	
	Currently, many reinforcement learning methods have been proposed, the most classic in which is Q-learning algorithm~\cite{b3-2, b3-3, b3-4}. The traditional Q-Learning method used in reinforcement learning is to create a Q value table to store state-action values. However, when the action space and state space are too large, it is not suitable for the Q value table to store. On the one hand, the computer's memory is limited, which will be the bottlenecks of storage performance. On the other hand, searching for the corresponding state in a large table is a time-consuming task. DQN can solve the above problems well~\cite{b3-5}. However, the target Q value of DQN is directly obtained by greedy policy. Selecting the largest Q value can make the Q value close to the possible optimization target, but it is easy to lead to overestimation. Therefore, in this paper, we use Double Deep Q-Network (DDQN) to realize our idea, since DDQN can reduce overestimation by decomposing the max operation in the target into action selection and action evaluation~\cite{b3-6}.
	
\begin{figure}
		\includegraphics[width=\linewidth]{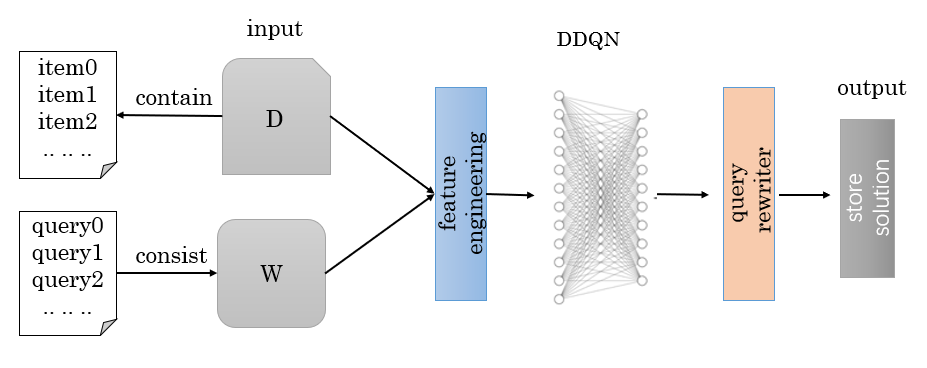}
		\centering
		\caption{System architecture of data storage}\label{3}
	\end{figure}

	The main system architecture of this article is shown in Figure~\ref{3}. First, the RDF graph data and workload are taken as the input. However, the data-scale is extremely large. We extract useful features and encode them as the input vector of the DDQN. In the training process, SQL query rewriting is required to ensure validity of workload. Finally, the optimized storage scheme is formed after the model training is completed.
		
 	To formulate the optimization problem of relational storage structure for graph data as a reinforcement learning problem, we need to formalize the generating process of storage structure as a Markov Decision Process (MDP). We will introduce the specific definition of important factors in this optimization method and give corresponding examples in Section~\ref{subsec:States},~\ref{subsec:Actions},~\ref{subsec:Rewards}, respectively. The Section~\ref{subsec:Predicate vector mapping} describes the feature extraction and vector mapping methods of data. In Section~\ref{subsec:Double Deep Q-Network Deep Learning Network}, the implementation of reinforcement learning network is introduced.

	\subsection{States}\label{subsec:States}
	$S$ represents a finite state set. For the state $S_t$ at any time $t$, it satisfies $S_t \in S$. $S_t$ represents the division of the table in the database at the current time. The initial state is $S_0$, and there is only one initial table $t_0$, which stores all data in the form of triples, indicating that the data has not been divided.
	
	Next, we illustrate the state transition process:
	
	\begin{itemize}
		\item The initial state $S_0 = \{t_0\}$, which means that there is only one initial table $t_0$ in the table set.
		
		\item If at time $T_1$, the reinforcement learning algorithm chooses the action to \textsf{divide} the table, and the condition is to divide the table according to $p= p_1$, then select the record of $p = p_1$ from the initial table $t_0$ and store it in a newly created triple table. The state at this time can be represented as $S_1 = \{t_0,t_1\}$.
		
		\item If at time $T_2$, the reinforcement learning algorithm chooses action to \textsf{divide} the table, and the condition is to divide the table according to $p = p_2$, then select the record of $p = p_2$ from the initial table $t_0$ and store it in a new triple table. It can be represented as $S_2 = \{t_0,t_1,t_2\}$.

		\item If at time $T_3$, the tables $t_1$ and $t_2$ \textsf{merge} according to a join condition, and the connection condition is $t_1.s=t_2.o$, the database will store the join results of $t_1$ and $t_2$ in a new table called $t_3$, the state at this time can be expressed as $S_3= \{t_0,t_1,t_2,t_3\}$.
		
		\item By analogy, if at the end of the final algorithm, the generated tables are finally $t_0, t_1, t_2, t_3$, and the storage structure of each table is different, then the data storage table scheme is the determined storage scheme.
	\end{itemize}
	
	When characterizing the state definition described above, we face two challenges. One is how to represent a triple pattern as a vector, and how to use the encoding to reflect the frequently queried sub-graph structure in these triple patterns. The other is that the vectors corresponding to the table sequence states are different at different time. For reinforcement learning algorithms, the dimensionality of the input data (the feature vector of the state) needs to be kept a constant, so it is more difficult to convert the triple pattern set of various numbers into a vector with a fixed length as the input of the network.
	
	In order to solve the above two problems, we map the triple table to the predicates it contains. This is because for a triple pattern, its predicates can best reflect the characteristics of the triple pattern, which is the most important information of a triple pattern. Therefore, it is appropriate to use a predicate to uniquely represent the triple pattern.
	
	After obtaining the result of the mapping, we encode and vectorize the predicate feature, and add a separation bit encoding to represent the separation between the two tables. However, this still cannot solve the problem of inconsistent vector dimensions obtained after different table partitioning states are characterized. In order to solve this problem, we need to predetermine the vector dimension of the characterization result when the data storage environment is initialized. After vectorizing all the predicate features, if the length of the vector obtained is less than the fixed dimension, then zeros are added at the end. Thus, the characterization result of each state can be turned into a fixed-length vector, which is convenient for normalizing the input of the neural network.
	
	\subsection{Actions}\label{subsec:Actions}
	Let $A$ represent a finite set of actions. For the state $A_t$ at any time $t$, it satisfies $A_t \in A$. The actions in the graph data storage process include actions of \textsf{dividing} and \textsf{merging}. \textsf{Dividing} action divides a table according to the predicate $p$, and the records of $p=p_1$ are selected from the initial table $t_0$ and stored in the newly created table. The table \textsf{merging} actions are divided into four categories, which are mainly based on the connection conditions of the two tables. It is denoted as $(t_i,t_j,flag)$. According to the flag condition, the triples $(s,p,o)$ in the tables $t_i$ and $t_j$ are combined. In the case that both $t_i$ and $t_j$ are obtained from the \textsf{dividing} table, $flag$ has four options:
	
	\begin{itemize}
		\item $flag = 1$, which means that the table is \textsf{merged} according to $t_i.s= t_j.s$;
		\item $flag = 2$, which means that the table is \textsf{merged} according to $t_i.s= t_j.o$;
		\item $flag = 3$, which means that the table is \textsf{merged} according to $t_i.o= t_j.s$;
		\item $flag = 4$, which means that the table is \textsf{merged} according to $t_i.o= t_j.o$;
	\end{itemize}
	
	However, if $t_i$ and $t_j$ may be obtained by the action of \textsf{merging} tables, then there will be more options for flag. Assuming that $t_i$ is \textsf{merged} from $m$ tables, and $t_j$ is \textsf{merged} from $n$ tables, the options for $flag$ total $4\times m \times n$ kinds.
	
\begin{figure}
		\includegraphics[width=\linewidth]{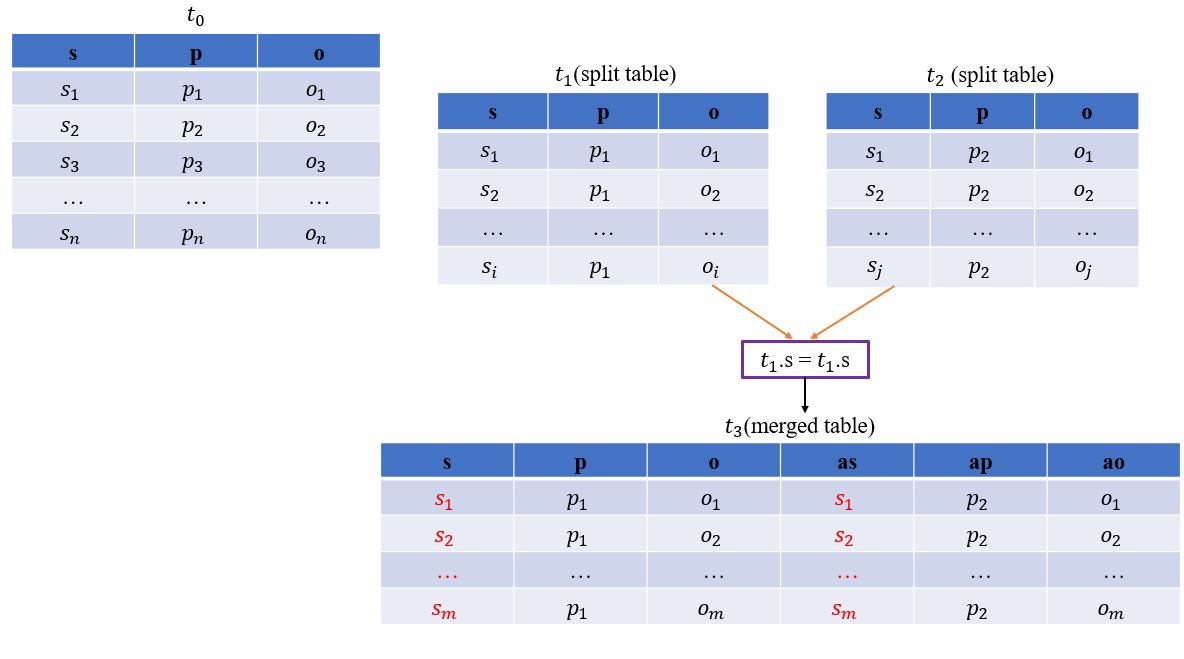}
		\centering
		\caption{The Actions of \textsf{dividing} and \textsf{merging} tables}\label{2}
	\end{figure}

	Take Figure \ref{2} as an example to illustrate the execution process of \textsf{dividing} and \textsf{merging} actions. The initial table $t_0$ contains $n$ predicates $\{p_1, \cdots ,p_n\}$, and there may be thousands of other records. With the table \textsf{dividing} action, the record of the predicate $p=p_1$ from $t_0$ is selected and stored in $t_1$; with the table \textsf{dividing} action, select the record of the predicate $p=p_2$ from $t_0$ is selected and stored in $t_2$; with the table \textsf{merging} action. The action is denoted as $(t_1,t_2,1)$, meaning that tables $t_1$ and $t_2$ are joined according to $t_1.s= t_2.s$, and the obtained records are stored in table $t_3$.
	
	\subsection{Rewards}\label{subsec:Rewards}
	Reward is the feedback information given by the environment after the learner performs an action. It is used to inform the learner of the result of the action and help the learner learn the value of the action. In the graph storage problem, the database executes the query of the workload according to the current table status at every step, and the performance of the data storage is measured by the query time. The less query time is used to execute the workload, the better the storage solution in this state is. Therefore, we use the increase in workload execution time compared to the previous state, i.e. the decreased time, as the reward in the current state. The more time the current state takes compared to the previous state query, the more efficient the query is, the greater the reward is, and the more effective the current action is.
	
	\subsection{Predicate vector mapping}\label{subsec:Predicate vector mapping}
	Since the scale of the graph data is extremely large, it is unrealistic to use all the data information as the input of the neural network. Therefore, our goal is to extract useful features of graph data and map them into fixed-length vector.
	
	A useful feature in graph data is its edge, which is the $p$ in the triple $(s,p,o)$, which can represent some information of the graph very well. Therefore, we vectorize the predicates in the graph data to represent the storage state of the current tables, and use the vector as the input of the neural network.
	
\begin{figure}
		\includegraphics[width=\linewidth]{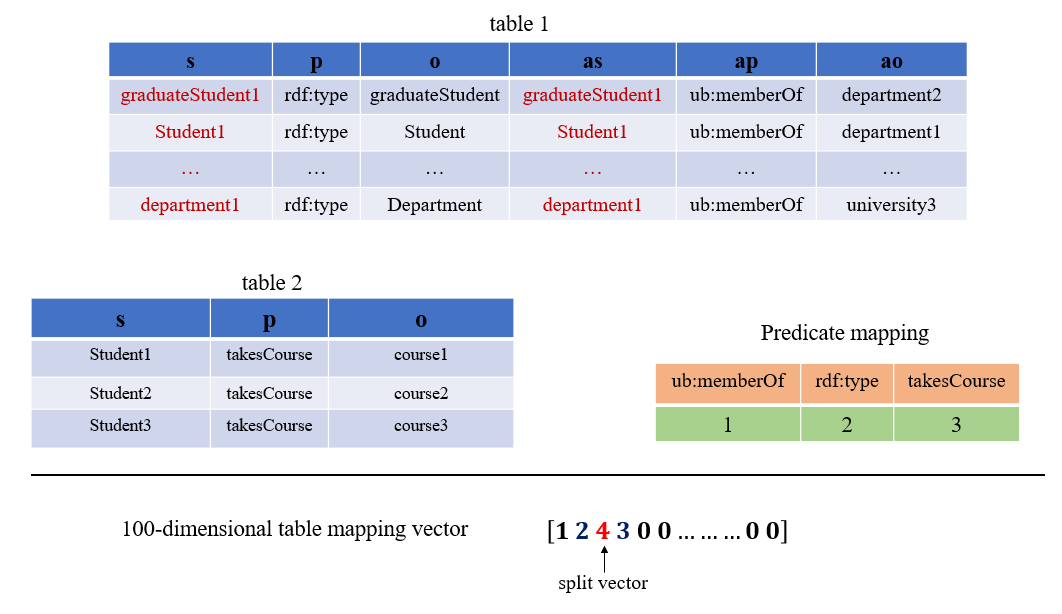}
		\centering
		\caption{Table state mapping process}\label{4}
	\end{figure}

	First, we count all the predicates in the graph, get the dictionary of $p$, and map each word to an integer from $1$ to $n$ ($n$ represents the total number of predicates). Next, we perform the mapping of the table state, and use the mapped vector as the input of the neural network below. An example of the table state mapping is shown in Figure~\ref{4}.
	
	Since the input vector dimension of the neural network is fixed, based on experience, we set the input vector dimension to 100 dimensions. Initially, the vector is all-zero. After a certain number of iterations, in the current state, the state space mapping method of the table is as follows:
	
	First, get all predicates in order of each table (except the initial table $t_0$) in the table sequence and perform vector mapping. After mapping the predicates of a table, an interval bit between different tables is added, and usually set to the maximum encoding of all the current predicates plus one. Next, the predicate is mapped on all other tables according to the above process. If the vector does not reach the dimension of the input vector, it is filled with zeros until it reaches the fixed input dimension.
	
	\subsection{Double Deep Q-Network Deep Learning Network}\label{subsec:Double Deep Q-Network Deep Learning Network}
	Since in reinforcement learning, our observational data is ordered, step by step, using such data to update the parameters of the neural network in DQN will cause overestimation problem. In supervised learning, \textsf{recall} depends on the data. Therefore, DQN uses the experience pool method~\cite{b3-7} to break the correlation between data and accelerate the convergence of the algorithm.
	
	We use DDQN neural network to solve the problem of overestimation by decoupling the target Q value action selection and target Q value calculation\cite{b3-8}. In the following, the specific implementation are introduced.
	
\begin{figure}
		\includegraphics[width=\linewidth]{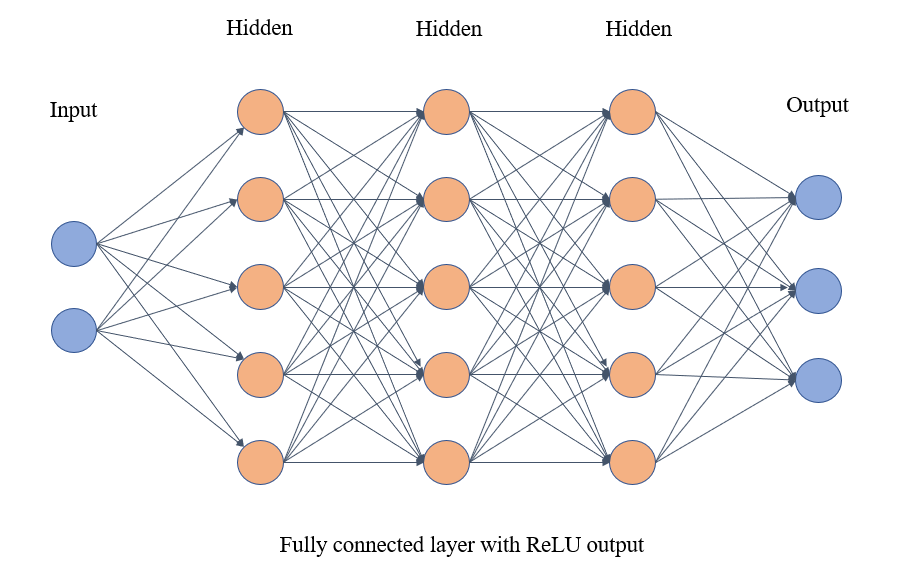}
		\centering
		\caption{Reinforcement learning network structure}\label{5}
	\end{figure}

	DDQN is composed of two identical neural network structures, namely the target network and the prediction network. The network architecture of the system is shown in Figure~\ref{5}. In order to improve the nonlinear mapping capability of the network, we add three hidden layers. Since the number of input dimensions is often smaller than the output dimensions, we first raise the dimensions of hidden layers to a value higher than the output dimension. We set two or more layers to increase the dimension gradually. Even if a deeper network may achieve higher accuracy, for efficiency issues, we design a three-layer fully connected layer hidden network.
	
	In the training process of the network, the DDQN neural network model chooses whether to \textsf{divide} or to \textsf{merge} in the current state. We set each round of training to a number of steps, and get the specific step with the largest accumulated reward, which should be the optimal step, as the reward feedback in each round. Therefore, this step and the previous reward feedback are more meaningful. Add them to the experience pool. During each training, the experience is randomly selected from the experience pool as the current network training sample.

	\section{Query rewrite}\label{sec:Query rewrite}
	SPARQL as a query language for RDF data, was released by W3C in 2008, has thus been a core focus of research and development for Semantic Web technologies~\cite{b4-1}. Since SPARQL queries cannot be used directly for query in relational database, a SPARQL-to-SQL module is required~\cite{b4-2}.

	Whether in the process of model training or using the trained model to decide the storage structure, query rewrite policy is indispensable. In the process of model training, the storage structure of the tables may change. Thus, using the same query statement may not conform to the optimal query policy. We will introduce the \textit{optimal rewrite policy} with examples in Section~\ref{sec:rewrite-policy}.	

	Query rewriting policy has an important impact on query efficiency. The reason is that if a poor policy choose a non-optimal one of all possible rewritten queries, the reward will be reduced. Thus, the agent will be misled. To ensure the optimality, we propose a priority-based rewriting policy to assist the generation of rewriting decisions. It can effectively solve the problem of how to select the optimal rewriting statement. This will be introduced in \ref{subsec:Priority list}.

	\subsection{Rewrite policy}\label{sec:rewrite-policy}
	As an important part of the system, query rewrite policy provides an important basis for reward in MDP. The query rewrite policy aims to rewrite the execution plan of the workload, i.e. the query statement, under a given state, to minimize the execution time. The rewrite policy is defined formally as follows:

	\begin{definition}
		Given the table sequence T: $t_1,t_2,\cdots,t_i,\cdots,t_n$, rewrite policy ${\Pi}$: \{raw query statement\}${\times}$\{rewritten query statement\}, for ${\forall s \in}$ \{raw query statement\}, if ${\Pi(s)}$ has the shortest execution time compared with other policies, it is the optimal one.
	\end{definition}

	For example, the following table sequence presents the predicates contained in the table and what join condition used to join:

	\begin{example}
		${t_0}$, $t_1$\{type\}, $t_2$\{comment\}, $t_3$\{topic\}, $t_4$\{type, comment, type.s=comment.s, comment.s=type.s\}, $t_5$\{type, comment, topic, type.s=comment.s, comment.s=type.s, comment.o=topic.s, topic.s=comment.o\}
		\\The raw SQL: \textit{select a.s from $t_0$ a, $t_0$ b, $t_0$ c where a.p = 'type' and a.o = 'senior user' and b.p = 'comment' and c.p = 'topic' and c.o = 'university' and a.s = b.s and b.o = c.s;}
		\\There may be many kinds of rewritten query statement, such as \textit{select a.s from $t_0$ a, $t_2$, $t_3$ where a.p='type' and a.o='senior user' and $t_3$.o = 'university' and a.s = $t_2$.s and $t_2$.o = $t_3$.s;} which uses $t_2$ and $t_3$ for query optimization.
		\\Some policy can also choose $t_4$ and $t_3$ for query optimization. In this case, the rewritten query statement should be \textit{select a.s from $t_4$, $t_3$ where $t_4$.o='senior user' and $t_3$.o = 'university' and $t_4$.bo = $t_3$.s;}.
		\\The optimal query rewritten policy $\Pi$ can give a rewritten query which has the shortest execution time.
	\end{example}

 	Since the policy $\Pi$ can give the optimal query rewriting under a given table sequence, the environment reward calculating will become more accurate, which has a more accurate guiding significance for the agent to update the action selection policy.

	\subsection{Priority list} \label{subsec:Priority list}
	Since query rewrite policy is of great significance to the calculation of reward value, it directly affects the feedback from the environment to the agent, thus affecting the agent's action selection policy. Therefore, the query rewrite policy should be as close to the facts as possible. First, from the empirical analysis of SQL execution time, we guess that the join of tables is the main factor of query time cost. In this way, the speed for a query executed in a joined table is faster than that in several single tables due to the cost of join. That is, the less the join times are, the faster the query execution speed is. But such experience cannot decide which query is faster when they all have the same join times.

	However, in the actual experiment (in Section \ref{subsec:Experimental results on execution time for priority items}), we discover that in the query execution of SQL, it is not always that the less the join times is, the faster the query speed is. Sometimes, the query execution time of SQL that needs twice joins is close to that of SQL that needs no join. Sometimes, the query execution speed of SQL with one join is slower than that of SQL with twice joins. The query execution time of two SQL query statements with the same join times may also differ greatly. Therefore, we can not judge the execution speed of rewritten SQL simply according to the join times.

	In the test, we find that the execution time of the rewritten SQL overturned our empirical conjecture, so we decide to use the real execution time of the rewritten queries of workload, i.e., to choose with practice experience as the query rewrite policy. We use exhaustive method to calculate the execution time of each rewritten SQL, sort by time, and generate our final priority item list. The algorithm is as follows:

	Algorithm \ref{alg:table-sequence-generation} takes the raw SQL $s$ as input and generates a table sequence which may be used for any rewritten query in the following process. Lines 1-3 creates all \textsf{single-table}s (tables which have one single type of predicate) from $t0$ by all predicates. Lines 4-6 creates all \textsf{merged} tables, which needs only one join, from the existing \textsf{single-table}s by any possible join condition. Lines 7-18 using traversal method, creates all \textsf{merged} tables which needs at least two joins. Let $k$ and $n$ denotes the number of join conditions and the maximum join times, respectively. The time complexity of Algorithm \ref{alg:priority-item-enumeration} should be $\Omega(nk^2)$.

	\begin{algorithm}
		\caption{Table Sequence Generation}
		\label{alg:table-sequence-generation}
		\scriptsize
		\begin{algorithmic}[1]
			\REQUIRE ~~\\
			Raw SQL $s$\\
			\ENSURE ~~\\
			All tables that may be used to execute any equivalent rewritten SQL queries in database.
			\FORALL {$pcond$ in all predicates involved in $s$}
			\STATE Create a new table with data divided by ${pcond}$ from t0;
			\ENDFOR
			\FORALL {$cond$ in all join conditions involved in $s$}
			\STATE Create a new table with data merged by ${cond}$;
			\ENDFOR
			\STATE ${n} \leftarrow$  the maximum join times;
			\FOR{$i \leftarrow 2$ to $n$}
			\STATE tableSet = \{all tables which needs $i-1$ times join\};
			\FOR{each ${t \in}$ tableSet}
			\FOR{each ${cond \in}$ all join conditions involved in ${s}$}
			\IF{$cond \notin$ join conditions involved in $s$ $\&\&$ $cond$ can be used on $t$ for join $\&\&$ the result of join with $cond$ on $t$ does not exist}
			\STATE $t\_$ $\leftarrow$ the table involved in $cond$ but not involved in $t$;
			\STATE Create a new table with data merged by $t$ and $t\_$ with $cond$;
			\ENDIF
			\ENDFOR
			\ENDFOR
			\ENDFOR
		\end{algorithmic}
	\end{algorithm}

	Algorithm \ref{alg:priority-item-enumeration} takes the table sequence generated by Algorithm \ref{alg:table-sequence-generation}, and outputs $tableSequence$ and $rewrittenSql$. This is the main algorithm for the traversal generation of priority items. Lines 3-36 uses depth-first search (DFS) with a stack storing the serial number of table. Let weight($t_i$) represents the weight of table $t_i$. The weight of a \textsf{single-table} is set to be the number of the table and the weight of a \textsf{merged} table is set to be the sum of all tables involved. $weightSum$ denotes the sum of all \textsf{single-table}s' weight. When $listWeight$, which is the weight of all tables in the stack, equals $weightSum$, the tables involved in any table in the stack exactly cover the whole dataset. Therefore, in Line 19, if the condition is true, it means that the tables in the stack are exactly involved in a possible rewritten query. Besides, if the condition in Line 23 is true, let $basicPriItem$ keep the priority item, so that Lines 37-39 can add priority items which may include $t0$ into $priSuccessTables$. When the weight of tables in the stack is greater than or equal to $weightSum$, the program will run to Lines 27-34. The table in stack will be popped up one at a time, and update related variables. And for loops from the next position of the stack popping position to the number of tables, Lines 8-22 are repeated to find all possible priority item using DFS. Let $n$ denote the number of tables in the table sequence. Let $k_{max}$ and $k_{min}$ be the maximum and minimum possible join times in any rewritten query, respectively. Apparently, $k_{max}$ should be less than the number of predicates in raw query. The time complexity of Algorithm~\ref{alg:priority-item-enumeration} should between $O(n^{k_{min}})$ and $O(n^{k_{max}})$.

	\begin{algorithm}
		\caption{Priority Item Enumeration}
		\label{alg:priority-item-enumeration}
		\scriptsize
		\begin{algorithmic}[1]
			\REQUIRE ~~\\
			the table sequence generated by Algorithm \ref{alg:table-sequence-generation}
			\ENSURE ~~\\
			$tableSequence$, the set of tables involved in the priority item\\
			$rewrittenSql$, the rewritten SQL query statement for the priority item
			\STATE $priSuccessTables$ $\Leftarrow$ $\emptyset$;
			\STATE $n$ $\leftarrow$ the number of tables in the table sequence;
			\WHILE {$i \textless n$}
			\STATE $stack \leftarrow \emptyset$, $stack$.push($i$);
			\STATE $listWeight$ $\leftarrow$ weight($t_i$)
			\STATE $tnameSet \leftarrow$ \{\textsf{single-table} name  including ${t_i}$\};
			\FOR{$j \leftarrow i+1$ to $n$}
			\STATE $tjTnameSet \leftarrow$\{all \textsf{single-table}s involved in $t_j$\}
			\IF{$listWeight+$weight($t_j$) $\leq$ $weightSum$ $\&\&$ $tjTnameSet \cap tnameSet = \emptyset$}
			\STATE $listWeight \leftarrow listWeight + $weight($t_j$);
			\STATE $stack$.push($j$);
			\STATE $tnameSet=tnameSet \cup tjTnameSet$;
			\ELSIF{$tjTnameSet \cap tnameSet \neq \emptyset$}
			\STATE continue;
			\ELSE
			\STATE break;
			\ENDIF
			\ENDFOR
			\IF{$listWeight=weightSum$}
			\STATE $successTable \leftarrow$ \{all tables in $stack$\};
			\STATE $rewrittenSql \leftarrow$ rewrite raw query so that it can be executed with tables in $successTable$;
			\STATE $priSuccessTable$.add(($successTable$, $rewrittenSql$));
			\IF{$successTable$ is the set of all \textsf{single-table}s}
			\STATE $basicPriItem\leftarrow$($successTable$, $rewrittenSql$);
			\ENDIF
			\ENDIF
			\WHILE{$stack \neq \emptyset$ }
			\STATE remove every \textsf{single-table} involved in the table on the top of the $stack$ from $tnameSet$;
			\STATE $listWeight \leftarrow listWeight - $ weight(the table on the top of the $stack$);
			\STATE $stack$.pop()
			\FOR{$j \leftarrow$ length($stack$)$+2$ to $n$}
			\STATE repeat 8-22;
			\ENDFOR
			\ENDWHILE
			\STATE $i \leftarrow i + 1$
			\ENDWHILE
			\IF{$basicPriItem$ is not null}
			\STATE Add priority item which may include table $t_0$ into $priSuccessTables$.
			\ENDIF
			\RETURN $priSuccessTables$
		\end{algorithmic}
	\end{algorithm}

	Algorithm \ref{alg:priority-list-generation} takes $priSuccessTable$ generated by Algorithm \ref{alg:priority-item-enumeration} as input, which contains $tableSequence$ and $rewrittenSql$. It calculates how much time each priority item costs and construct join conditions from the rewritten SQL. Line 1-2 sort the priority items by the (rewritten) query execution time. Line 3-6 construct join conditions and store them into $chooseInfo$. This is the basis of satisfaction when searching the table sequence in rewriting query. Line 5 checks the connectivity of join conditions i.e. if there are join condition that $t_i$.s=$t_j.s$ and $t_j$.s=$t_k$.s, $t_i$.s=$t_k$.s should also be in this join condition. Let $n$ be the size of priority list. The time complexity of Algorithm \ref{alg:priority-list-generation} should be $O(n)$.

	\begin{algorithm}
		\caption{Priority List Generation}
		\label{alg:priority-list-generation}
		\scriptsize
		\begin{algorithmic}[1]
			\REQUIRE ~~\\
			$priSuccessTable$ generated by Algorithm \ref{alg:priority-item-enumeration}
			\ENSURE ~~\\
			$tableSequence$, ditto\\
			$rewrittenSql$, ditto\\
			$executeTime$, the average of three times of the rewritten SQL execution time\\
			$chooseInfo$, keeps the sufficient condition for the priority item
			\STATE Execute SQL query in the database three times for each priority item in $priSuccessTable$. Record execution time and get the average.
			\STATE Sort $priSuccessTable$ by the average time of execution time.
			\FOR{each $result$ in $priSuccessTable$}
			\STATE Construct join condition as $chooseInfo$ from the rewritten SQL;
			\STATE Check connectivity of join conditions in $chooseInfo$;
			\STATE Add $chooseInfo$ into $priSuccessTable$;
			\ENDFOR
			\RETURN $priSuccessTable$
		\end{algorithmic}
	\end{algorithm}
		
	\section{Experimental results}\label{sec:Experimental results}
	The experimental environment is Huawei Cloud Server. The CPU is Intel(R) Xeon(R) Gold 6151 CPU @ 3.00GHz. The memory is 64G memory. The system is CentOS Linux release 7.4.1708 (Core), and the database environment is MySQL 8.0.21.
		
	The datasets used in this experiment are WatDiv~\cite{b5-1} dataset and LUBM~\cite{b5-2} dataset.
	\begin{enumerate}
		\item WatDiv data. The data set is designed for measuring how an RDF data management system performs across a wide spectrum of SPARQL queries with varying structural characteristics and selectivity classes. It provides data generator and query templates. We choose 10 different scale datasets with scale factor from 2 to 20 with the interval of 2. Dataset at scale factor 1 contains 105,257 triples, 5,597 distinct subjects, 85 distinct predicates, 13,258 distinct objects. We execute 20 queries on the data.
		
		\item LUBM data. The data set consists of a university domain ontology, customizable and repeatable synthetic data. We generate a data set with 850 universities from based on it. The generated data contains 117,489,648 triples, 27,941,697 distinct subjects and objects, and 18 distinct predicates. Since the LUBM dataset is very large, we only selected two thousandths of the scale as the test data, so the original workload is slightly modified to make it possible to have query results on small datasets. We execute 10 relatively simple queries on them.				
	\end{enumerate}

	We test our approach on these two datasets on our model, which will be introduced in Section \ref{subsec:Self-result}. We also conduct a comparative experiment on Apache Jena to verify the performance of the model, which will be introduced in Section \ref{subsec:Compare with Jena SDB}. In Section \ref{subsec:Experimental results on execution time for priority items}, experimental results about empirical conjecture are showed. Summary of experimental results is discussed in Section \ref{subsec:Summary}. 

		\subsection{Self-result}\label{subsec:Self-result}
		\begin{table}[htbp]
		\caption{LUBM dataset test results}\label{table:LUBM data set test results}
		\begin{center}
			\begin{tabular}{ccccc}
				\hline
				Episodes & t1$^{\mathrm{1}}$ (s) & t2$^{\mathrm{2}}$ (s) & increase & space(M) \\
				\hline
				5 & 3.355 & 0.392 & 88.32\% & 420.43 \\
				10 & 3.367 & 0.386 & 88.54\% & 420.43 \\
				20	& 3.359 & 0.358 & 89.34\% & 510.72 \\
				30	& 3.371 & 0.341 & 89.88\% & 510.55 \\
				40	& 3.341 & 0.387 & 88.42\% & 420.43 \\
				\hline
				\multicolumn{4}{l}{$^{\mathrm{1}}$t1 is the original time.} \\
				\multicolumn{4}{l}{$^{\mathrm{2}}$t2 is the time after promotion.}
			\end{tabular}
		\end{center}
	\end{table}
		
		As can be observed from Table \ref{table:LUBM data set test results}, the execution time of the original LUBM workload fluctuates between 3.34s and 3.37s. With the increase in the number of training episodes, the optimized query time (t2) is continuously decreasing, and the improvement percentage is constantly increasing. At round 30, the improvement percentage has reached 89.88\%.
		
		\begin{table}[htbp]
		\caption{WatDiv data set test results}\label{table:WatDiv data set test results}
		\begin{center}
			\begin{tabular}{ccccc}
				\hline
				Episodes & t1$^{\mathrm{1}}$ (s) & t2$^{\mathrm{2}}$ (s) & increase & Space(M) \\
				\hline
				5 & 7.586 & 2.686 & 64.59\% & 99.07 \\
				10 & 7.569 & 2.714 & 64.14\% & 102.21 \\
				15 & 7.568 & 0.606 & 91.99\% & 296.62 \\
				20 & 7.563 & 0.519 & 93.14\% & 295.63 \\
				25 & 7.57 & 0.521 & 93.12\% & 295.63 \\
				30 & 7.559 & 0.522 & 93.09\% & 295.63 \\
				35 & 7.571 & 0.521 & 93.12\% & 294.64 \\
				50 & 7.552 & 0.514 & 93.19\% & 295.38 \\
				100 & 7.571 & 0.518 & 93.16\% & 295.38 \\
				150 & 7.577 & 0.521 & 93.12\% & 295.38 \\		
				\hline
				\multicolumn{4}{l}{$^{\mathrm{1}}$t1 is the original time.} \\
				\multicolumn{4}{l}{$^{\mathrm{2}}$t2 is the time after promotion.}
			\end{tabular}
			\label{tab1}
		\end{center}
	\end{table}
		
		From Table \ref{table:WatDiv data set test results}, the execution time of the original WatDiv workload is less than 8s. With the increase in the number of training episodes, the optimized query time is continuously decreasing, and the improvement percentage is constantly increasing. At round 20, the improvement percentage has reached 93.12\%.
		
		\begin{figure}[htbp]
		\includegraphics[width=\linewidth]{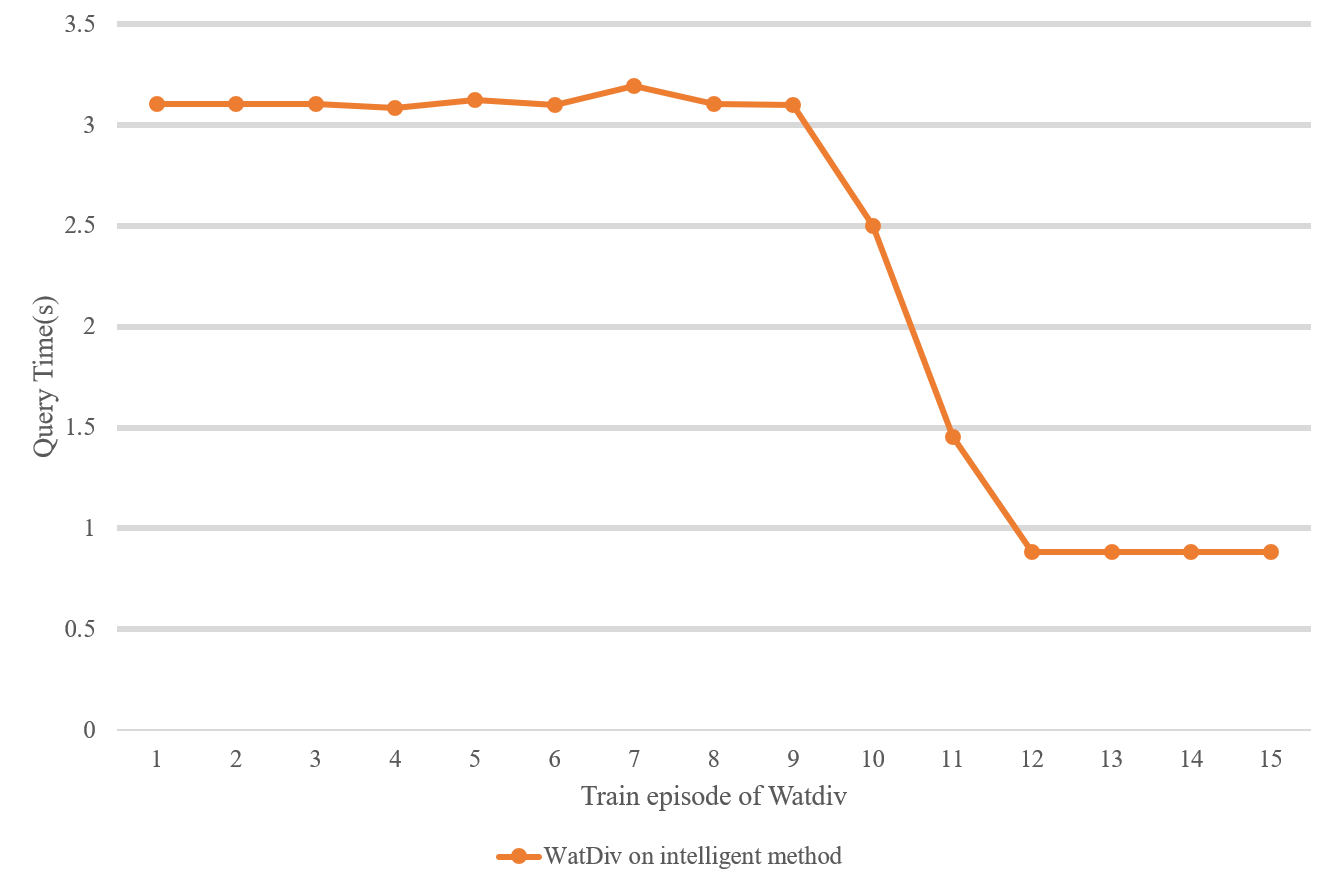}
		\centering
		\caption{Query time changes with the number of training episodes}\label{fig:Query time changes with the number of training episodes}
	\end{figure}
		
		Figure \ref{fig:Query time changes with the number of training episodes} shows the convergence of optimization with the increase of episodes. Before the 9th episode, the query time has not been significantly optimized. After the 9th episode, the optimized query time gradually becomes shorter. From the 12th episode, the optimization effect begins to converge. It shows the convergence of the reinforcement learning method.
				
		The results above show that the reinforcement learning method can effectively improve the query efficiency.
		
		\subsection{Compare with Jena SDB}\label{subsec:Compare with Jena SDB}
		Apache Jena (or Jena in short)\cite{b5-3} is a free and open source Java framework for building Semantic Web and Linked Data applications. The framework is composed of different APIs interacting together to process RDF data. Jena has two data storage methods. One is TDB, which is the native storage method of Jena. The other is SDB, which is a persistent triple stores method using relational databases. SDB uses an SQL database for the storage and query of RDF data. In this experiment, we use MySQL~\cite{b5-4} as the backend storage database for SDB.
		\begin{table}[htbp]
			\caption{Experimental results of WatDiv data on Jena}\label{table:Test results of WatDiv data on Jena}
			\begin{center}
				\begin{tabular}{ccccc}
					\hline
					Datasets &
					space1$^{\mathrm{1}}$ (M) &
					space2$^{\mathrm{2}}$ (M) &
					t1$^{\mathrm{3}}$ (s) & t2$^{\mathrm{4}}$ (s) \\
					\hline
					watdiv2 & 30.67 & 45.22 & 34.695 & 5.383 \\
					watdiv4 & 60.56 & 90.34 & 70.064 & 9.746 \\
					watdiv6 & 92.51 & 134.38 & 110.694 & 15.156 \\
					watdiv8 & 123.24 & 165.56 & 154.925 & 26.481 \\
					watdiv10 & 152.63 & 204.66 & 197.395 & 33.424 \\
					watdiv12 & 184.88 & 258.72 & 251.377 & 31.635 \\
					watdiv14 & 216.33 & 287.77 & 305.563 & 40.178 \\
					watdiv16 & 245.83 & 329.03 & 361.361 & 40.635 \\
					watdiv18 & 278.11 & 390.09 & 420.701 & 55.420 \\
					watdiv20 & 307.43 & 424.17 & 486.605 & 57.562 \\		
					\hline
					\multicolumn{4}{l}{$^{\mathrm{1}}$space1 is the dataset size on disk.} \\
					\multicolumn{4}{l}{$^{\mathrm{2}}$space2 is the storage space size on database.} \\
					\multicolumn{4}{l}{$^{\mathrm{3}}$t1 is the total time of construction.} \\
					\multicolumn{4}{l}{$^{\mathrm{4}}$t2 is the total time of queries in workload.}
				\end{tabular}
				\label{tab1}
			\end{center}
		\end{table}	
	
		According to Table \ref{table:Test results of WatDiv data on Jena}, the storage space of the tables created by Jena is small. However, in the process of the experiment, we find that there are empty tables created by Jena. In some degree, this demonstrates that the data storage on Jena is unevenly distributed.
		
\begin{figure}[htbp]
			\includegraphics[width=\linewidth]{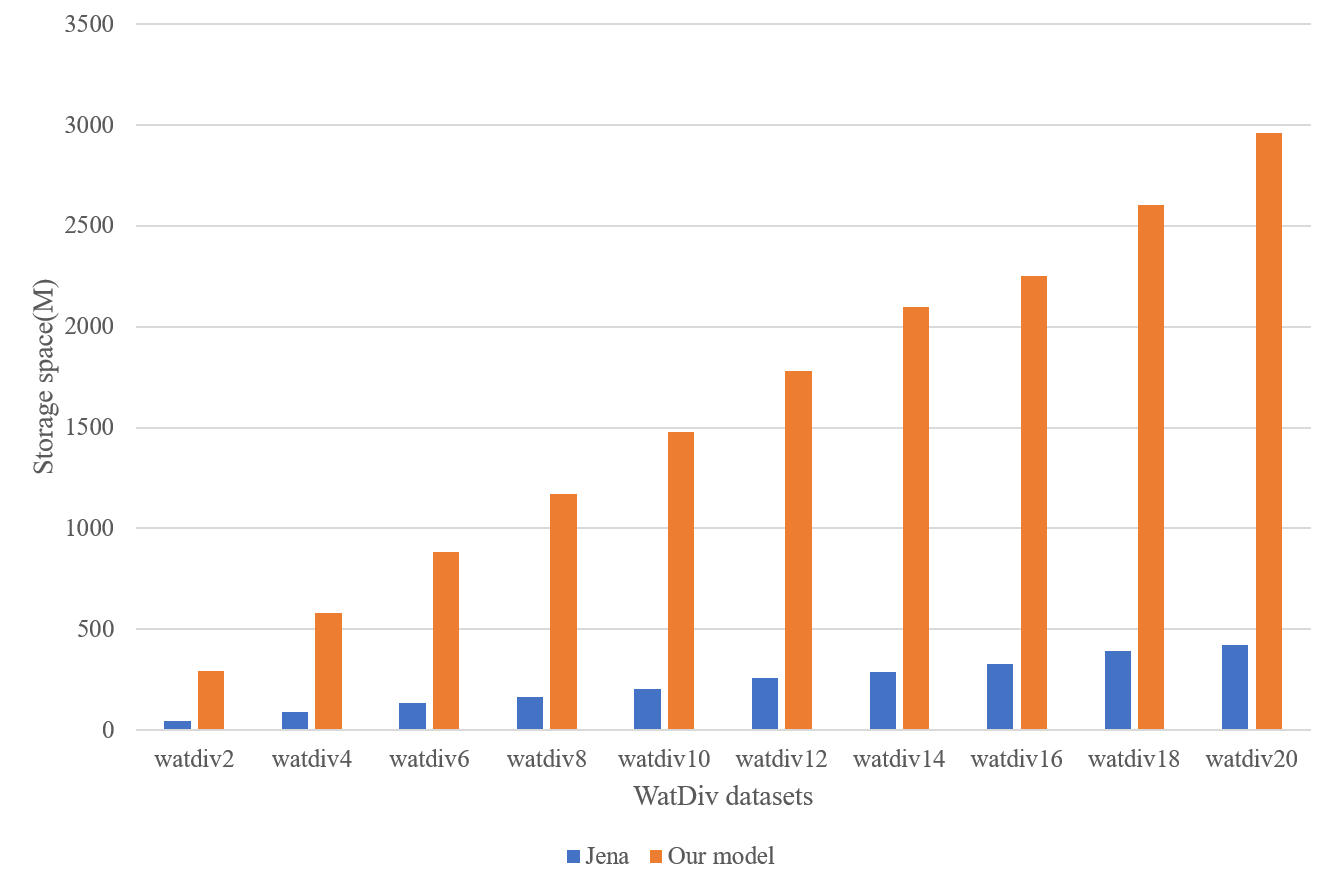}
			\centering
			\caption{The storage space of WatDiv datasets on Jena and our approach}\label{fig:The storage space of WatDiv datasets on Jena and our model}
		\end{figure}

\begin{figure}[htbp]
			\includegraphics[width=\linewidth]{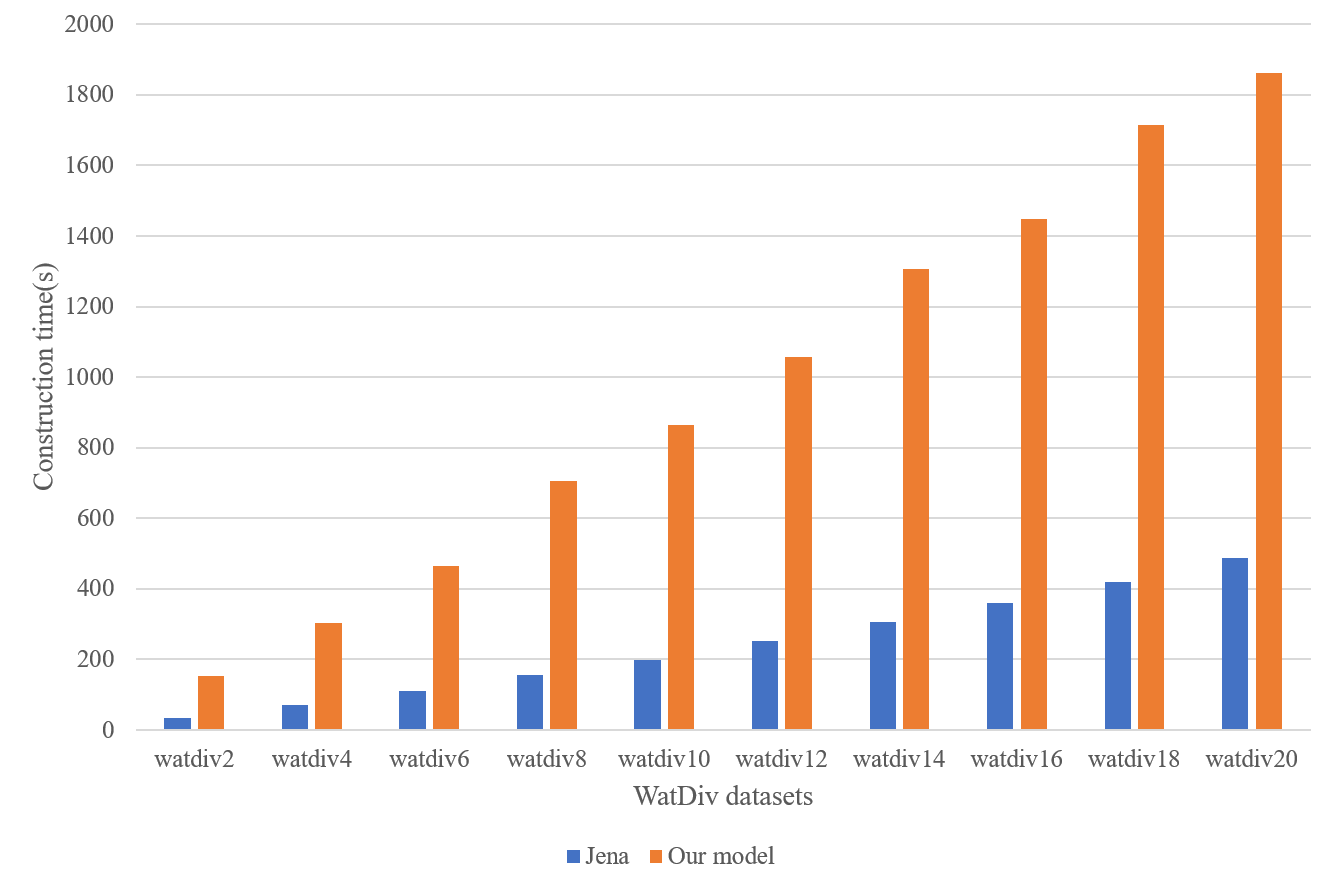}
			\centering
			\caption{The time of construct tables on Jena and our approach}\label{fig:The time of construct tables on Jena and our model}
		\end{figure}

\begin{figure}[htbp]
			\includegraphics[width=\linewidth]{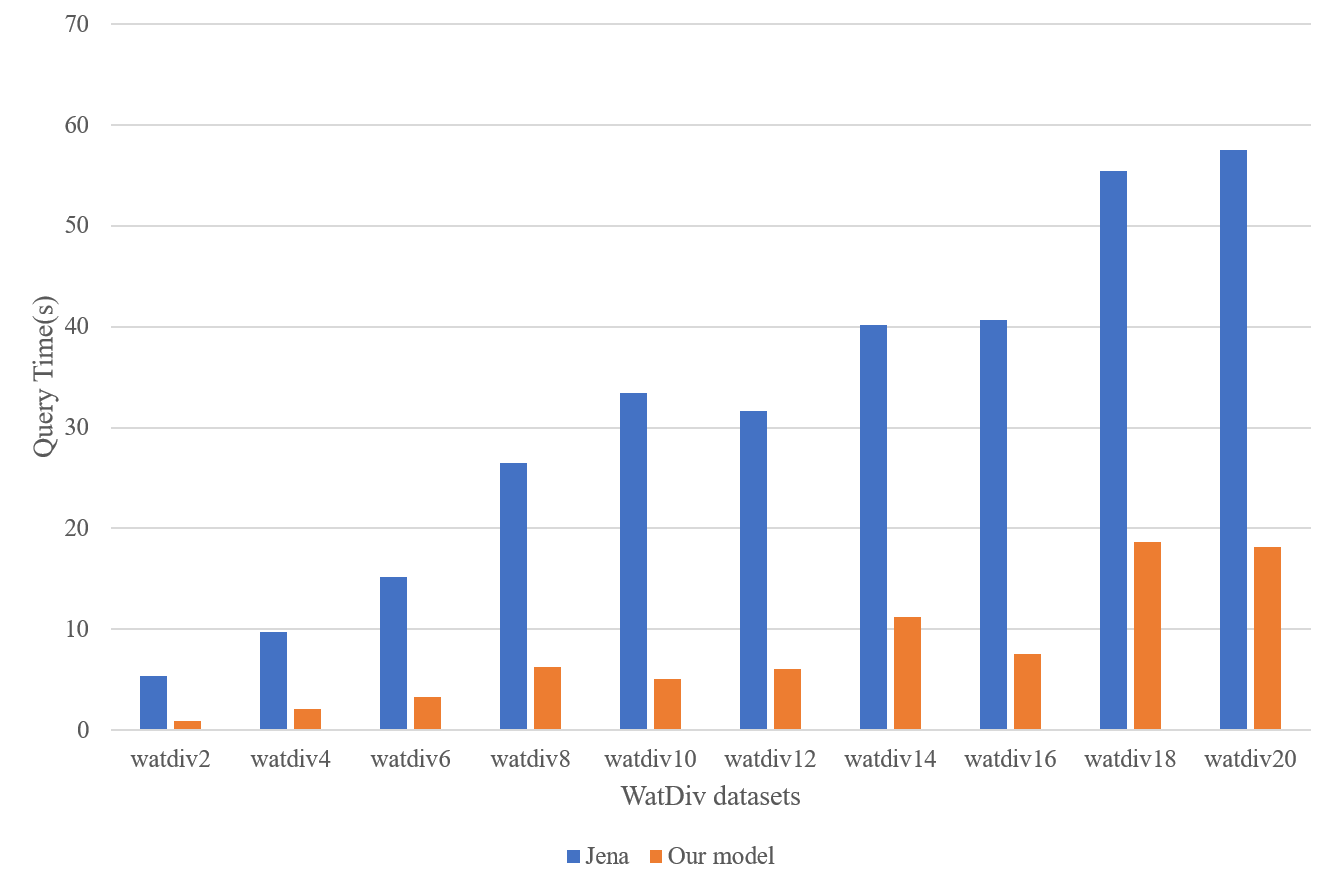}
			\centering
			\caption{The query time of WatDiv datasets on Jena and our apporach}\label{fig:The query time of WatDiv datasets on Jena and our model}
		\end{figure}

		We test the performance on various data sizes and observe how the performance of Jena and our approach changes as the increase of data scale. We can see that the storage spaces on our model are larger than on Jena in Figure \ref{fig:The storage space of WatDiv datasets on Jena and our model}. And the time of table construction is longer than Jena according to Figure \ref{fig:The time of construct tables on Jena and our model}. However, the query time is much less than Jena according to Figure~\ref{fig:The query time of WatDiv datasets on Jena and our model}.	
	
\begin{figure}[h]
			\includegraphics[width=\linewidth]{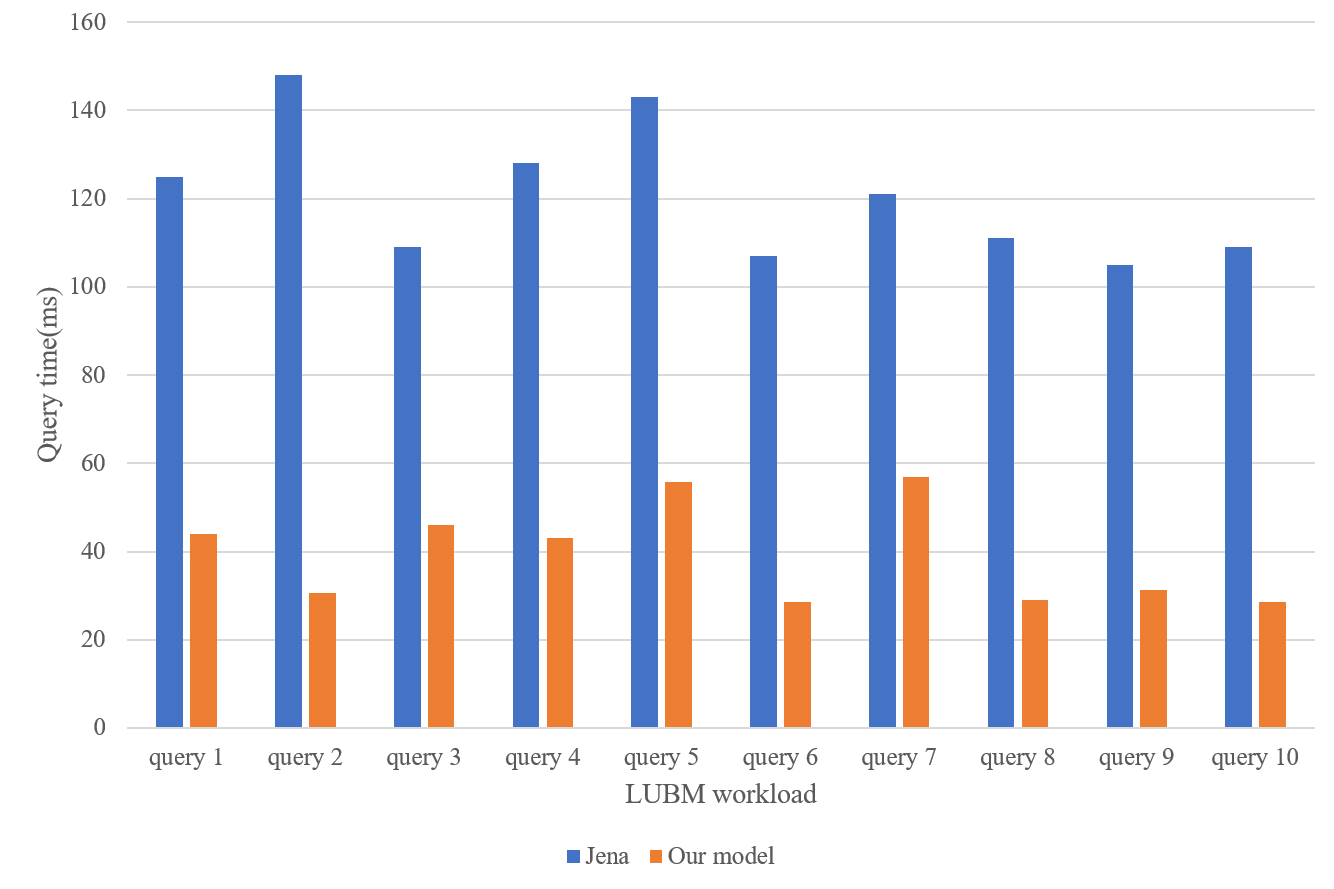}
			\centering
			\caption{The query time of LUBM workload on Jena and our approach}\label{fig:The query time of LUBM workload on Jena and our model}
		\end{figure}

		We also test the LUBM dataset on Jena, and the experiment results compared with our model are shown in Figure~\ref{fig:The query time of LUBM workload on Jena and our model}.	We run 10 queries respectively to compare Jena and our approach. We observe that the query time on our approach is less than 60ms. However, all query time on Jena exceeds 100ms. Especially on query2, the time on our approach reduces by 117.3ms compared to Jena. From the results, our storage structure significantly improves query performance.	
		
		\subsection{Execution time for priority items}\label{subsec:Experimental results on execution time for priority items}
		During the study on how to construct \textsf{priority list} for rewriting query, we find that the empirical conjecture which holds that the less the join times are, the faster the query execution speed is, is not always true. This is confirmed in Example~\ref{exp:2}.
	
		\begin{example}\label{exp:2}
		In the \textsf{priority list} of \textit{select a.o, b.s from t0 a, t0 b, t0 c where a.o = c.o and b.s = c.s and a.s = 'http://db.uwaterloo.ca/\textasciitilde galuc/wsdbm/City0' and a.p = 'http://www.geonames.org/ontology\#parentCountry' and b.p = 'http://db.uwaterloo.ca/\textasciitilde galuc/wsdbm/likes' and b.o = 'http://db.uwaterloo.ca/\textasciitilde galuc/wsdbm/Product0' and c.p = 'http://schema.org/nationality'}, a few items are showed as followed, which overturns out empirical conjecture.
		\begin{enumerate}
			\item Let $tp0$ denote the table divided by \textit{'http://www.geonames.org/ontology\#parentCountry'} from $t0$. Let $tp3$ denote the table merged by the \textsf{single-table} $a$ with predicate \textit{'http://schema.org/nationality'} and the \textsf{single-table} $b$ with predicate \textit{'http://db.uwaterloo.ca/\textasciitilde galuc/wsdbm/likes'} with the condition $a.s=b.s$. Thus, the rewritten query should make use of $tp0$ and $tp3$ for execution, \textbf{which has one single join. The execution time is 0.659ms.} The rewriten query is \textit{select tp0.o, tp3.s from tp0,tp3 where tp0.o=tp3.bo and tp0.s='http://db.uwaterloo.ca/\textasciitilde galuc/wsdbm/City0' and tp3.o='http://db.uwaterloo.ca/\textasciitilde galuc/wsdbm/Product0'}.

\item Let $tp0$ denote the table divided by \textit{'http://www.geonames.org/ontology\#parentCountry'} from $t0$. Let $tp1$ denote the table divided by the \textit{'http://db.uwaterloo.ca/\textasciitilde galuc/wsdbm/likes'}. Let $tp2$ denote the table divided by the \textit{'http://schema.org/nationality'}. Thus, the rewritten query should make use of $tp0$, $tp1$ and $tp2$ for execution, \textbf{which has two joins. The execution time is 2.227ms.} The rewritten query is \textit{select tp0.o, tp1.s from tp0,tp1,tp2 where tp1.s=tp2.s and tp0.o=tp2.o and tp0.s='http://db.uwaterloo.ca/\textasciitilde galuc/wsdbm/City0' and tp1.o='http://db.uwaterloo.ca/\textasciitilde galuc/wsdbm/Product0'}.

			\item Let $tp5$ denote the table merged by the \textsf{single-table} $a$ with predicate \textit{'http://www.geonames.org/ontology\#parentCountry'}, the \textsf{single-table} $b$ with the predicate \textit{'http://db.uwaterloo.ca/\textasciitilde galuc/wsdbm/likes'} and the \textsf{single-table} $c$ with predicate \textit{'http://schema.org/nationality'}, with the condition $a.o=c.o$ and $b.s=c.s$. Thus, the rewritten query should make only use of $tp5$ for execution, \textbf{which has no join. The execution time is 10.267ms.} The rewritten query is \textit{select tp5.o, tp5.bs from tp5 where tp5.s='http://db.uwaterloo.ca/\textasciitilde galuc/wsdbm/City0' and tp5.bo='http://db.uwaterloo.ca/\textasciitilde galuc/wsdbm/Product0'}.
		\end{enumerate}
	\end{example}

		\subsection{Summary}\label{subsec:Summary}
		Overall, our approach can converge in considerable time. The increase of execution time after convergence can reach to 89\% (LUBM) and 93\% (WatDiv), with the cost of about 7 times extra storage space. At different scales of dataset, the improvement also performs well. In comparison to WatDiv, even though our storage space is about 6 times of Jena, and the construction time is also longer, our query time is one-sixth to one-third of Jena. For each query in LUBM workload, our execution time is about one-third to one-fourth of Jena. Therefore, the storage structure constructed by our method can effectively improve query performance.
		
	\section{Conclusion}\label{sec:Conclusion}		
	To solve the problem of automatical relational storage structure design for graph data, we propose an approach based on reinforcement learning. The proposed method comprehensively uses the characteristics of data and workload. At the same time, the method is universal, i.e. there is an optimal storage solution for different graph datasets, which can greatly improve the data storage efficiency under large datasets. Experimental results demonstrate that proposed approach could generate efficient storage structure. In future work, we plan to design a more complex and more effective storage structure for queries to make our approach more complete.

\end{document}